\documentclass[twocolumn,aps,prl,showpacs,preprintnumbers,amsmath,amssymb, superscriptaddress]{revtex4}

\usepackage{graphicx}
\usepackage{dcolumn}
\usepackage{bm}

\begin{document}


\title{Percolation through Voids around Randomly Oriented Faceted Inclusions} 


\author{D. J. Priour, Jr and N. J. McGuigan}
\affiliation{Department of Physics \& Astronomy, Youngstown State University, Youngstown, OH 44555, USA}


\date{\today}

\begin{abstract}
We give a geometrically exact treatment of percolation through voids around assemblies of 
randomly placed impermeable barrier particles, introducing a computationally 
inexpensive approach to finding critical barrier density thresholds marking the transition 
from bulk permeability to configurations which do not support fluid or charge transport in the 
thermodynamic limit.  We implement a dynamic exploration technique which accurately determines 
the percolation threshold, which we validate for the case of randomly placed spheres.  We find 
the threshold densities for randomly oriented hemispherical fragments and tablets with flat and 
curved surfaces derived from a sphere truncated above and below its equator.  To incorporate an  
orientational bias, we consider barrier particles with dipole moments along the symmetry axis; the 
extent of the alignment is then tuned with uniform electric fields of varying strengths. The latter 
compete with thermal fluctuations which would eliminate orientational bias in the absence of 
an applied field.
\end{abstract}

\maketitle

Percolation transitions, with all of the hallmarks of a continuous phase transition, are disorder mediated shifts from 
configurations allowing transport (e.g. fluid or charge) to systems not navigable in the bulk limit.
Scenarios in which neighboring or overlapping sites form connected clusters clusters are amenable to 
cluster infiltration tools such as the Hoshen Kopelman algorithm~\cite{Hoshen}
to determine if a configuration of disorder percolates, allowing transport along its full extent.
Void percolation, relevant to a variety of porous media, is in a sense complementary with transport taking place through     
irregularly shaped spaces surrounding impenetrable barrier particles instead of through the inclusions themselves. Salient cases include porous 
media supporting fluid or ion transport through spaces among impermeable grains, which range in shape from round weathered grains well     
approximated as spheres to angular crystallites with randomly oriented facets as an additional element of disorder.

In seeking the threshold density $\rho_{c}$ above which system spanning void networks are disrupted, analyses similar  
to the Hoshen Kopelman algorithm must in general be applied in conjunction with an approximate treatment, such as a discrete mesh   
superimposed on the irregular volumes between barrier particles~\cite{Martys,Maier,Yi1,Yi2}.  In special cases techniques such as 
Voronoi tesselation have been successfully applied~\cite{Elam,Marck,Rintoul}, but are difficult to generalize to non-spherical inclusions.
In cases compatible with Cartesian symmetry such as aligned cubes, the void geometry may be worked out~\cite{Koza}.
In this work, we calculate $\rho_{c}$  with an 
exact treatment of barrier particle geometry and a modest computational investment,  circumventing an 
explicit characterization of void shapes by using dynamical explorations of void networks with virtual tracer particles. 
The tracers propagate from the center of a large cubic sample of randomly placed inclusions, following linear 
paths punctuated by specular reflection from barrier surfaces as the tracer encounters impermeable inclusions.  
We consider the case of randomly placed spheres to validate a novel computationally efficient method for finding 
percolation thresholds around randomly placed barriers in 3D.  In addition, we examine randomly oriented hemispheres and tablets 
with the latter formed from spheres truncated above and below the equator.  To our knowledge this is the first calculation of percolation 
thresholds for randomly oriented faceted inclusions.  Moreover, we validate a technique using dynamical exploration to accurately 
determine if a barrier concentration $\rho$ is above or below the percolation threshold, thereby providing a simple  
tool for finding percolation transitions.

In dynamical explorations, one may determine the likelihood of a tracer particle traversing a specified distance or 
alternatively accumulate statistics 
on distances traveled during a fixed time.
Previous examples of dynamical simulations in 3D have been described in the context of randomly 
placed spheres~\cite{Kammerer,Hofling,Spanner}  
We present results from both finite volume simulations, with effectively infinite dwell time (i.e. with 
observables of interest, such as the mean fraction of tracers emerging from the simulation volume
 converged with respect to the allocated time), and finite time simulations with an effectively infinite volume in the 
sense that tracer particles never breach the system boundary. In the finite time scenario 
one is not impacted by computational latency inherent in finite volume calculations, where to ensure convergence of observables  
we use dwell times in excess of  mean escape times by at least an order of magnitude.
Using finite time simulations, we calculate critical quantities such as threshold concentrations $\rho_{c}$ of inclusions to within a few tenths of 
a percent with a computational investment equivalent of less than a week on a contemporary multi-core work station.

In both the finite volume and finite time calculations, we average over at least $10^{4}$ realizations of disorder, cubic 
volumes of randomly placed barriers with a predetermined concentration $\rho$.  Pertinent length scales such as the size of the cubic simulation volume 
and tracer particle traversal distances
are specified in terms of the sphere radius for spherical inclusions or the radius of the circumscribing sphere for hemispheres and 
tablets.  Dynamical explorations in the finite volume scenario typically occur in cubic configurations with 
edge lengths from $L = 20$ to 
$L = 60$ in the case of randomly placed spheres.   
However, for finite time calculations, larger 
simulation volumes, from $L = 200$ to  $L = 500$, are used 
to ensure none of the tracers escape while allowing for at least on the order of $10^{6}$ collisions.   
To achieve this, dwell times range from $5 \times 10^{5}$ for more dense 
assemblies of plate-like particles to $10^{7}$ 
time units (with normalized inter-collision velocities where $| \vec{v} |  = 1$) in the case of spherical barriers.
Tablets are parameterized with $\epsilon$, the distance above and below the equator where the sphere is truncated.
Smaller $\epsilon$ values correspond to thinner tablets, and we  consider the case $\epsilon = 0$ (infinitely thin plates) as well.

For the sake of computational efficiency in finding trajectories for tracer particles, cubic simulation volumes are 
partitioned into voxels a unit on a side, and for optimal memory usage we  
generate disorder realizations voxel by voxel.  The probability relation of $n$ 
inclusions in a voxel, $P(n) = (\rho v)^{n} e^{-\rho v}/n!$
($v = 1$ being the voxel volume), 
allows for the efficient and statistically accurate sampling of $P(n)$ with a robust random number generator.  
For faceted inclusions, a random orientation of the axis normal to the flat surface or surfaces
is chosen in an isotropic manner.  To identify
collisions, neighboring voxels are checked for penetration of spheres or circumscribing spheres in the case of 
non-spherical defects.  For the latter, a further determination is made if the trajectory contacts a facet or curved region.  
The shortest time determines the collision point, with specular reflection fixing the modified velocity.

While we report chiefly on finite time calculation results,  we briefly discuss finite volume calculations of critical indices, such 
as the correlation length critical exponent $\nu$, most readily accessed by sampling quantities such as the mean fraction $f$ of 
tracer particles
escaping in the finite volume context. The latter is 
determined by the connectivity of void networks, in turn governed by the correlation length $\xi$, which 
diverges as $\xi \sim (\rho - \rho_{c})^{-\nu}$ immediately below $\rho_{c}$.  Though we primarily seek $\nu$, other variables 
exhibit the generic scaling form $X \sim (\rho - \rho_{c} )^{\alpha_{x}}$ near $\rho_{c}$, and we calculate $\alpha_{f}$ for the fraction of 
escaping tracers to verify universal behavior for void percolation phenomena in 3D.
 
To calculate critical indices,  we appeal to single parameter finite size scaling theory, which posits a scaling form 
$X(\rho) = L^{\alpha_{X}/\nu} g[L^{1/\nu} (\rho - \rho_{c} )]$~\cite{Stauffer}  
near $\rho_{c}$. Accordingly, we plot
$L^{-\alpha_{f}/\nu}f$ with respect to $L^{1/\nu} (\rho - \rho_{c})$, with Monte Carlo data for various system sizes $L$ and $\rho$ in
principle lying on the same curve with a correct choice of $\rho_{c}$, $\nu$, and $\alpha$.  To use the data collapse 
phenomenon as a quantitative tool we consider  
a scaling function $g(x) = \sum_{j=0}^{n} A_{j} x^{j}$, optimizing with  
respect to the $A_{j}$ coefficients as well as $\rho_{c}$, $\nu$, and $\alpha_{f}$ via nonlinear least squares fitting.  The polynomial 
form for $f(x)$ with comparatively few coefficients is warranted by the approximately linear variation of $f(x)$ near $x = 0$. 
\begin{figure}
\includegraphics[width=.4\textwidth]{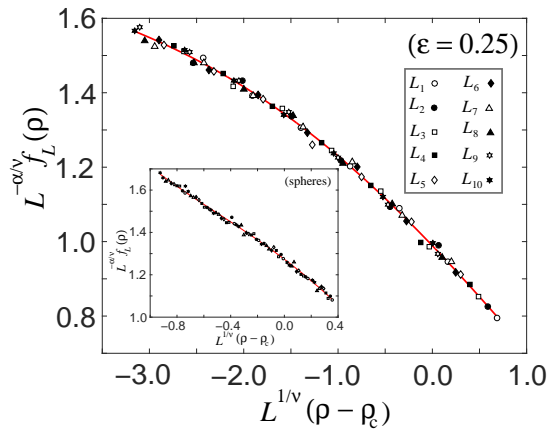}
\caption{\label{fig:Fig1} (Color online) Data Collapse plots for finite volume calculations for tablets with $\epsilon = 0.25$ and 
spheres in the inset. Symbols correspond to Monte Carlo data, and solid lines to analytical curves.  The legend 
applies to both the main and inset graphs, with system sizes ranging from $L_{1} = 18$ to $L_{10} = 32$ for the former and 
$L_{1} = 33$ to $L_{10} = 60$ for the latter.}
\end{figure}
Sample data collapses appear in Fig.~\ref{fig:Fig1} with tablets ($\epsilon = 0.25$) in the main graph and randomly placed spheres in the inset.  
In discussing void percolation phenomena, it is customary also to specify the excluded volume $\phi_{c} = e^{-\rho_{c} v_{\mathrm{B}} }$ with 
$v_{\mathrm{B}}$ the barrier volume; for spheres we find $\rho_{c} = 0.831(3) [\phi_{c} = 0.0308(5)]$ while for 
hemispheres we obtain $\rho_{c} = 1.47(2) [\phi_{c} = 0.046(3)]$. Calculated $\nu$ values, such as $\nu_{\mathrm{sphere}} = 0.91(10)$ and $\nu_{\mathrm{hemisphere}} = 0.81(25)$,
while crude,
are compatible with the 3D percolation exponent $\nu = 0.875$~\cite{Stauffer} for discrete percolation transitions.  The $\alpha_{f}$ exponents are each internally 
consistent within bounds of Monte Carlo error with, e.g., $\alpha_{f}^{\mathrm{sphere}} = -0.24(4)$ and $\alpha_{f}^{\mathrm{hemisphere}} = -0.26(12)$.
In terms of computational efficiency, these results are a significant improvement relative to a dynamical exploration study involving one of 
us~\cite{dynamic}.

Results of dynamical exploration  in the finite time scenario are also amenable to finite size scaling analysis
tools, but with respect to time instead of system size.  To this end, we divide the total dwell time for a tracer 
particle into ten evenly spaced deciles with statistics recorded at the conclusion of each decile in the context 
of a single simulation.  That a single simulation provides results for many times offers a significant compuational 
advantage.  For an appropriate scaling observable, we consider 
$\delta = (\langle x_{\mathrm{max}} - x_{\mathrm{min}} \rangle + \langle y_{\mathrm{max}} - y_{\mathrm{min}} \rangle  + 
\langle z_{\mathrm{max}} - z_{\mathrm{min}} \rangle )/3$
where $x_{\mathrm{max}}$ and $x_{\mathrm{min}}$ are maximum and minimum values, respectively attained for the $x$ 
coordinate with analogous relations for $y$ and $z$, with angular brackets indicating 
averages over disorder.  The $\delta$ quantity is by design an arithmatic mean of Cartesian coordinates instead  of 
the square root of a Euclidian sum to include another layer of averaging with fluctuations in one coordinate partially damped  by the other 
components.
\begin{figure}
\includegraphics[width=.39\textwidth]{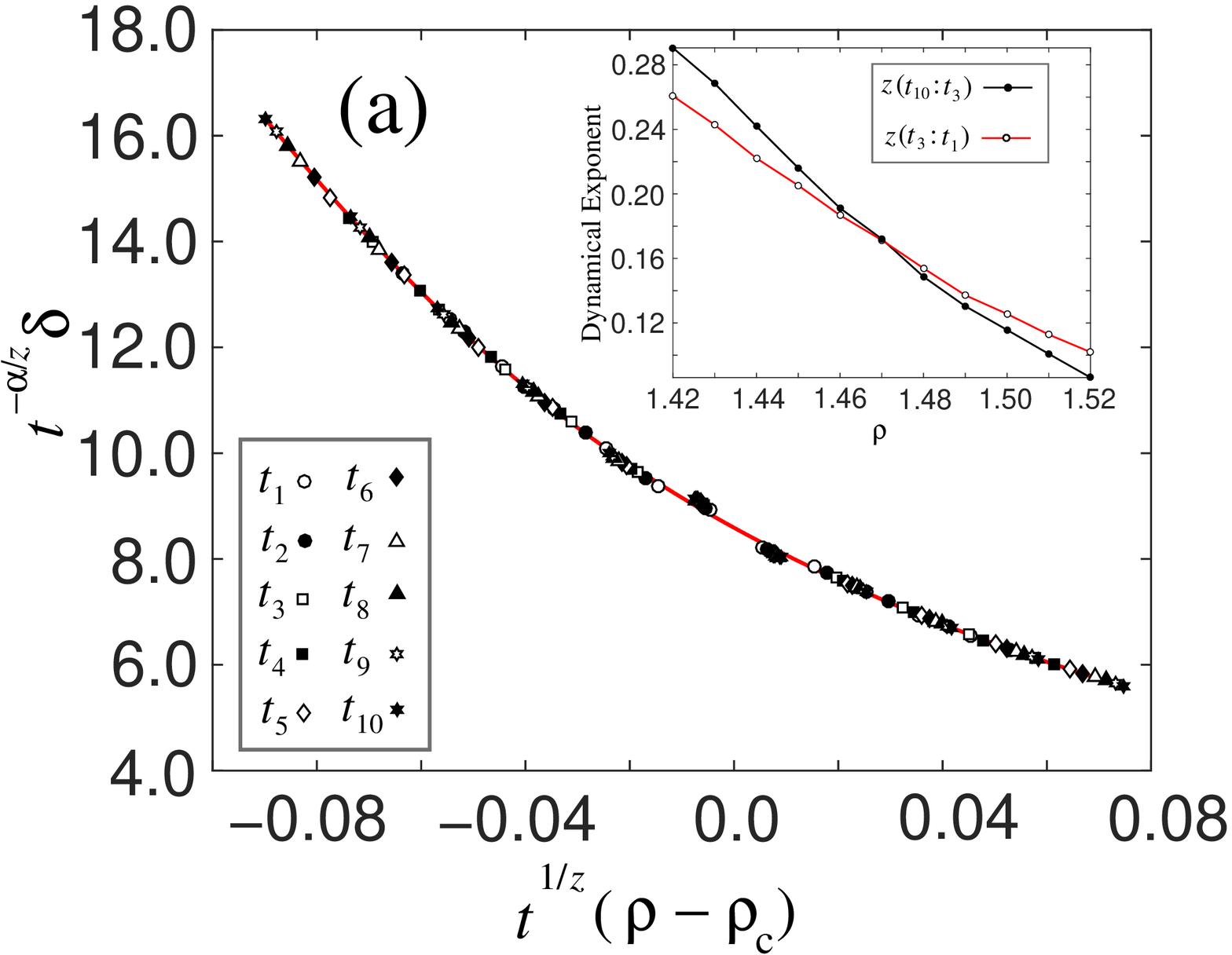}
\includegraphics[width=.37\textwidth]{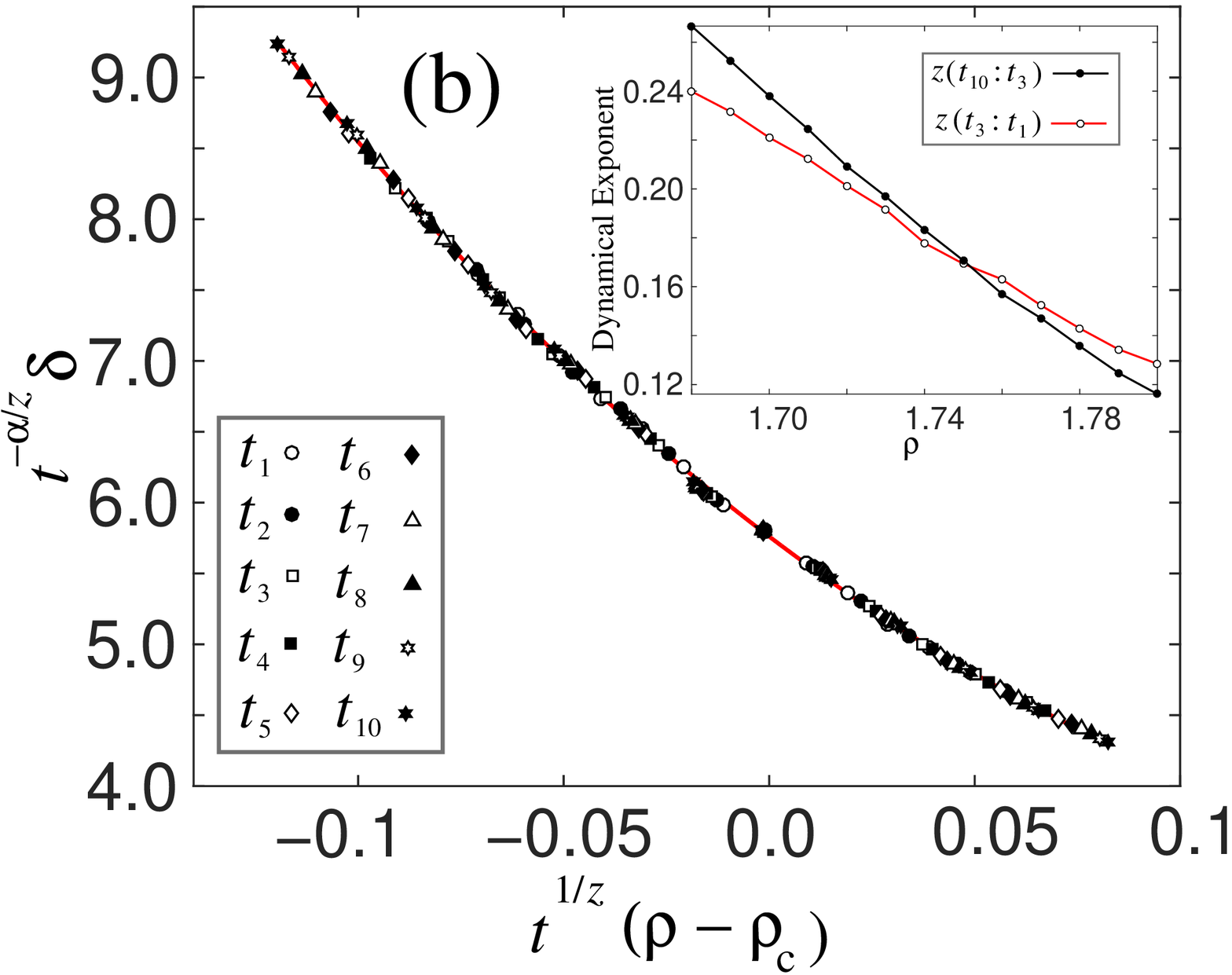}
\caption{\label{fig:Fig2} (Color online) Data Collapse plots for finite time calculations in the case of 
hemispheres (panel a), and tablets for $\epsilon = 0.25$ (panel b).  Symbols correspond to Monte Carlo data for deciles as indicated 
in the legend and solid lines to analytical curves.}
\end{figure}
Sample data collapse plots are shown in Fig.~\ref{fig:Fig2} for spheres in panel (a), hemispheres in panel (b), and 
tablets in the case $\epsilon = 0.25$ in panel (c), with symbols again representing Monte Carlo data and 
the solid line being the scaling function.  Data collapses yield the percolation threshold concentrations 
$\rho_{c}$ to within a few tenths of a percent and in the case of finite time calculations also serve to validate 
a simpler and comparably accurate alternative technique involving comparing dynamical scaling exponents across different time 
intervals.  For $t$ very large, one might expect an asymptotically power law dependence $\rho \sim t^{z}$ for $\rho < \rho_{c}$ 
or $\rho = \rho_{\infty}$ for $\rho > \rho_{c}$, tantamount to $z = 0$, where $z$ is a dynamical exponent. 
Diffusive transport with $z = 1/2$ would be expected for $\rho < \rho_{c}$.  Nevertheless, in practice
for finite times, apparent subdiffusive behavior could be manifest even below $\rho_{c}$, and a finite $z$  
discerned for $\rho > \rho_{c}$  with some tracer particles  in transit even though confined to finite void networks.  
Thus, one could in principle consider an effective dynamical exponent $z_{\mathrm{eff}}(t)$ which would increase 
to an asymptotic diffusive value of 1/2 for $\rho < \rho_{c}$ and decrease to zero for $\rho > \rho_{c}$.   
A  $z_{\mathrm{eff}}$  has previously been quantified and used in a 2D context~\cite{Schnyder}

We exploit this behavior to determine with a single simulation if the barrier concentration is above or below the percolation 
threshold by considering $z(t_{2}:t_{1}) \equiv \ln [\delta(t_{2})/\delta(t_{1})]/\ln (t_{2}/t_{1})$, an effective dynamical 
exponent for the interval from $t_{1}$ to $t_{2}$. Hence, for $z(t_{10},t_{3})$ and $z(t_{3},t_{1})$, one  would expect
$z(t_{10},t_{3}) < z(t_{3},t_{1})$ for $\rho < \rho_{c}$ with the inequality reversed for $\rho > \rho_{c}$ and thus a crossing of the curves 
for $\rho = \rho_{c}$.  The insets of the plots in Fig.~\ref{fig:Fig2} indicate $z(t_{10},t_{3})$ and $z(t_{3},t_{1})$ crossings at  
$\rho_{c}$.
Aside from computational convenience, effective exponent curve crossings yield percolation thresholds to a level of accuracy comparable to 
that of data collapses, and hence may be used in lieu of the latter to glean the critical concentration $\rho_{c}$.
\begin{figure}
\includegraphics[width=.4\textwidth]{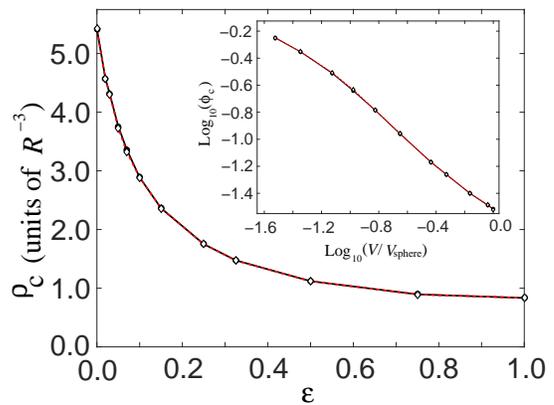}
\caption{\label{fig:Fig3} (Color online) Critical concentrations of tablets ranging from 
(spheres, $\epsilon = 1.0$) to plate-like inclusions ($\epsilon = 0$).   The inset is a log-log 
graph with the normalized volume on the horizontal axis and the excluded volume on the vertical axis. Filled symbols correspond to 
data collapse results and open symbols to exponent curve crossings.}
\end{figure}

In the main graph of Fig.~\ref{fig:Fig3},  percolation thresholds are plotted with respect to 
the tablet thickness parameter $\epsilon$.  
Results are in accord with finite volume calculations though with improved accuracy with, e.g., $\rho_{c}^{\textrm{spheres}} = 0.834(1) [\phi_{c} = 0.0304(1)]$
for spheres where $\epsilon = 1$.
The inset graph shows the critical excluded volume $\phi_{c}$ plotted with respect 
to the tablet volume $V_{\epsilon} = 2 \pi \epsilon (1 - \epsilon^{2}/3)$, excluding $\epsilon = 0$ (i.e. circular plates).  Solid symbols in both graphs    
represent data collapse results, while open circles indicate percolation thresholds obtained from effective exponent crossings.  A marked rise in the 
threshold concentration is evident with decreasing $\epsilon$; nevertheless, $\rho_{c}^{\textrm{plates}} = 5.42(6)$ is finite for  
circular plates, and in accord with $\rho_{c} = 5.457(8)$ from Y.~B.~Yi and K.~Esmail~\cite{Yi2}. 
In the log-log plot in the inset the $\phi_{c}$ curve, essentially linear for a decade with decreasing inclusion volume, begins to level off markedly for 
$\epsilon < 0.05$.   Not depicted are results for 
randomly oriented hemispheres, with $\rho_{c} = 1.475(6) [\phi_{c} = 0.0455(6)]$.
\begin{figure}
\includegraphics[width=.35\textwidth]{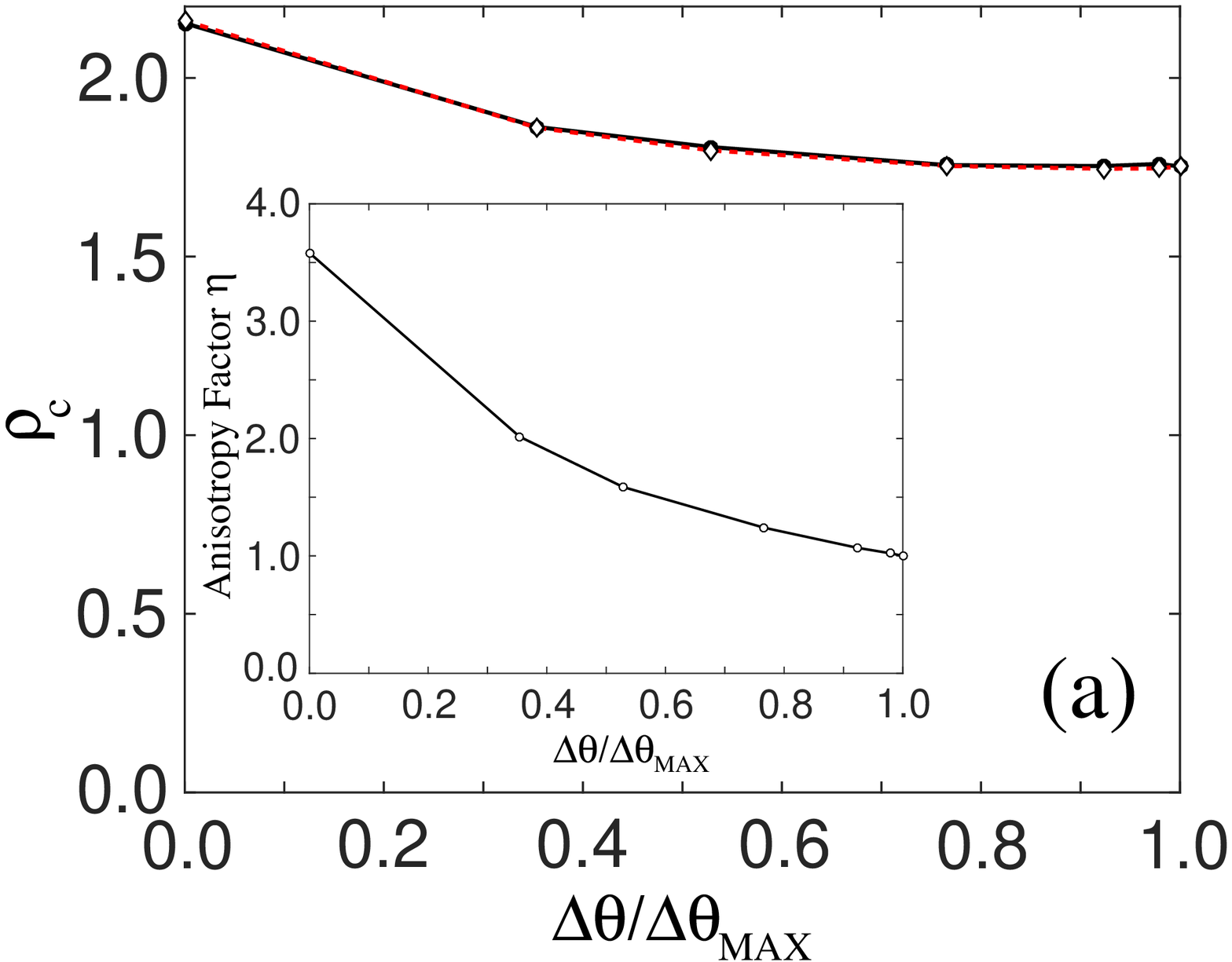}
\includegraphics[width=.35\textwidth]{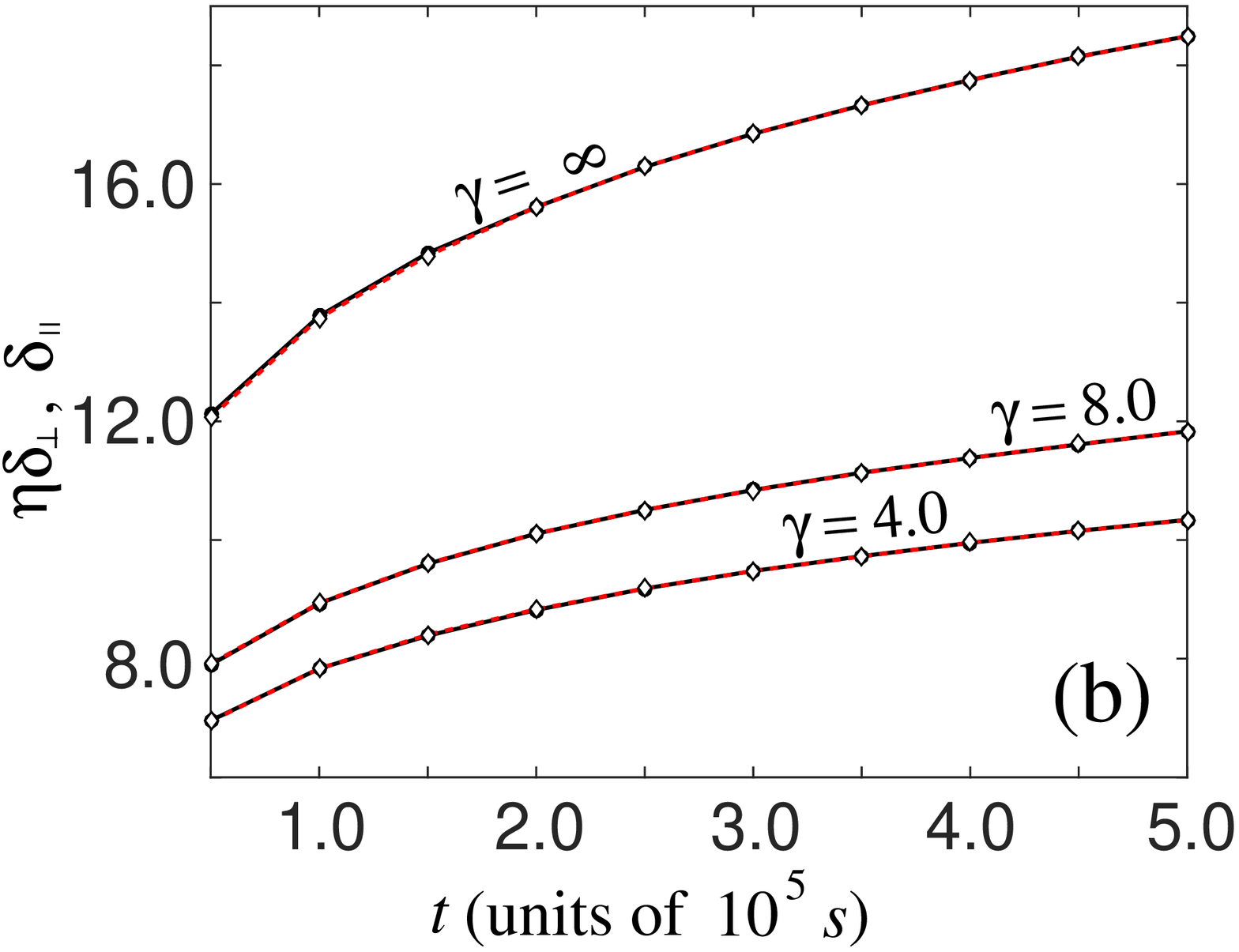}
\caption{\label{fig:Fig4} (Color online) Panel (a) is a graph of percolation thresholds with respect to the relative width of the angle 
distribution $\Delta \theta/\Delta \theta_{\mathrm{max}}$, while the inset shows the anisotropy factor $\eta$ with respect to 
$\Delta \theta/\Delta \theta_{\mathrm{max}}$ as well.  Filled and open symbols indicate $\rho_{c}$ values for perpendicular and 
parallel transport respectively. Panel (b) shows time dependent curves with filled and open symbols indicating the scaled 
perpendicular motion and parallel motion respectively.} 
\end{figure}

In this work, we have examined assemblies of faceted inclusions with no orientational order.  Nevertheless, relevant situations
may call for a partial or complete alignment of plate-like barriers to modify the percolation threshold and/or transport characteristics.  As
a viable physical mechanism to establish a tunable degree of alignment, we consider tablets imbued with an electric dipole moment $\vec{p}$ aligned with the
tablet normal axis, with the system as a whole subject to a uniform electric field $\vec{E}$.  With the dipole orientational energy $-\vec{p} \cdot \vec{E}$
in competition with thermal fluctuations, the relevant Boltzmann Factor is $e^{\gamma \cos \theta}$, a distribution we sample 
with a technique similar to the Heat Bath algorithm~\cite{Wocht}
in the context of the Heisenberg model for magnetism on 3D lattices.
The parameter $\gamma \equiv p E/(k_{\mathrm{B}} T)$ may be adjusted (i.e. by changing either the field strength or the temperature) to choose any
desired intermediate level of alignment. A salient question is whether aligned tablets lead to a distinct percolation transitions for flow perpendicular
versus flow parallel
to the tablet flat surfaces.  In fact, we find the transitions in both cases to be identical for completely  
and partially aligned tablets, evident in the main 
graph in Fig.~\ref{fig:Fig4} where $\rho_{c}$ is plotted versus $\Delta \theta/\Delta \theta_{\mathrm{max}}$, the standard deviation
of $\theta$ relative to $\Delta \theta_{\mathrm{max}}$, for $E = 0$, with no orientational bias.   

In a broad sense, the percolation threshold increases modestly with increasingly alignment.  However, despite slight variation in  $\rho_{c}$ 
with $\Delta \theta/\Delta \theta_{\mathrm{max}}$, reducing $\Delta \theta$ has a more significant impact on the dynamics parallel and perpendicular to 
the tablet planes.  Panel (b) shows superimposed time dependent $\delta_{\perp}$ and $\delta_{\parallel}$ curves 
near $\rho_{c}$ for several $\gamma$ values, where the $\delta_{\perp}$ 
curves are scaled by an anisotropy factor $\eta$, leading to close overlap with the $\delta_{\parallel}$ curves.  Notwithstanding the modest increase in 
$\rho_{c}$, for perfect alignment $\eta = 3.58$, implying  a significant suppression of transport perpendicular to the tablet plates.

In conclusion, using dynamical exploration simulations, we have calculated percolation thresholds for voids 
around faceted inclusions, including randomly oriented hemispheres and tablets with an exact treatment of the geometry 
of the barriers.  Though independent calculations in the fixed volume and fixed time frameworks yield identical results, the 
latter yield comparable accuracy with a smaller computational effort.   We have validated 
a simple technique for finding the critical concentration of barrier particles with modest computational demands which 
permits an immediate and accurate determination if a given $\rho$ is above or below $\rho_{c}$.
We have examined the effect of a partial or complete alignment of plate-like inclusions, finding a modest increase in 
$\rho_{c}$ even for $\Delta \theta = 0$, though with a significant enhancement of transport in directions parallel to the 
tablet planes. 

\begin{acknowledgments}
Calculations described here have benefitted from use of the OSC, the Ohio Supercomputer facility~\cite{OSC}
\end{acknowledgments}



\begin{thebibliography}{11}
\expandafter\ifx\csname natexlab\endcsname\relax\def\natexlab#1{#1}\fi
\expandafter\ifx\csname bibnamefont\endcsname\relax
  \def\bibnamefont#1{#1}\fi
\expandafter\ifx\csname bibfnamefont\endcsname\relax
  \def\bibfnamefont#1{#1}\fi
\expandafter\ifx\csname citenamefont\endcsname\relax
  \def\citenamefont#1{#1}\fi
\expandafter\ifx\csname url\endcsname\relax
  \def\url#1{\texttt{#1}}\fi
\expandafter\ifx\csname urlprefix\endcsname\relax\def\urlprefix{URL }\fi
\providecommand{\bibinfo}[2]{#2}
\providecommand{\eprint}[2][]{\url{#2}}


\bibitem{Hoshen} J.~Hoshen and R.~Kopelman, Phys. Rev. B, \textbf{14}, 3438 (1976).

\bibitem{Martys} N.~S.~Martys, S.~Torquato, and D.~P.~Bentz, Phys. Rev. E \textbf{50}, 403 (1994).

\bibitem{Maier} R.~S.~Maier, D.~N.~Kroll, H.~T.~Davis, and R.~S.~Bernard, J. Colloid Interface Sci. 
\textbf{217}, 341 (1999).

\bibitem{Yi1} Y.~B.~Yi, Phys. Rev. E \textbf{74}, 031112 (2006).

\bibitem{Yi2} Y.~B.~Yi and K.~Esmail, J. Appl. Phys. \textbf{111}, 124903 (2012).

\bibitem{Elam} W.~T.~Elam, A.~R.~Kerstein, and J.~J.~Rehr, Phys. Rev. Lett. \textbf{52}, 1516 (1984). 

\bibitem{Marck} S.~C.~van der Marck, Phys. Rev. Lett. 77, 1785 (1996).

\bibitem{Rintoul} M.~D.~Rintoul, Phys. Rev. E \textbf{62}, 68 (2000).

\bibitem{Koza} Z.~Koza, G.~Kondrat, and K.~Suszczy\'{n}ski, J. Stat. Mech.: Th. Exp. \textbf{11}, P11005 (2014).

\bibitem{Kammerer} A.~Kammerer, F.~H\"{o}fling, and T.~Franosch, EPL \textbf{84}, 66002 (2008). 

\bibitem{Hofling} F.~H\"{o}fling, T.~Munk, E.~Frey, and T.~Franosch, J. Chem. Phys. \textbf{128}, 164517 (2008).
 
\bibitem{Spanner} M.~Spanner, F.~H\"{o}fling, G.~E.~Schr\"{o}der-Turk, K.~Mecke, and T.~Franosch,
J. Phys.: Condens. Matter \textbf{23}, 234120 (2011).

\bibitem{Stauffer} D.~Staffer and A.~Aharony, \textit{Introduction to Percolation Theory},
2nd ed. (Taylor and Francis, Bristol, UK, 1992). 

\bibitem{dynamic} D.~J.~Priour, Jr., Phys. Rev. E, \textbf{89}, 012148 (2014). 

\bibitem{Schnyder} S.~K.~Schnyder,  M.~Spanner, F.~H\"{o}fling, T.~Franosch. and J.~Horbach,
Soft Matter \textbf{11}, 701 (2015).

\bibitem{Wocht} F.~R.~Brown and T.~J.~Woch, Phys. Rev. Lett., \textbf{58}, 2394 (1987). 

\bibitem{OSC} Ohio Supercomputer Center.  1987.  Ohio Supercomputer Center.  Columbus, OH:  Ohio
Supercomputer Center.  http://osc.edu/ark:/19495/f5s1ph73.

\end{thebibliography}
\end{document}